\documentclass[preprint,aps,pra]{revtex4}
\usepackage{graphicx,amsmath,natbib}

\begin{document}

\title{Combustion of biomass as a global carbon sink}

\author{Rowena Ball}
\email{Rowena.Ball@anu.edu.au}
 \homepage{www.rsphysse.anu.edu.au/~rxb105/rb.html } 
\affiliation{
Department of Theoretical Physics \& Mathematical Sciences Institute\\ The Australian National University,
 Canberra 0200 Australia}

\date{April 09, 2008}

\begin{abstract}
This note is intended to highlight the important role of black carbon produced from biomass burning in the global carbon cycle, and encourage further research in this area. Consideration of the fundamental physical chemistry of cellulose thermal decomposition suggests that suppression of biomass burning or biasing burning practices to produce soot-free flames must inevitably transfer more carbon to the atmosphere. A simple order-of-magnitude quantitative analysis indicates that black carbon may be a significant carbon reservoir that persists over geological time scales.

\end{abstract} 
\maketitle
A recent review article by Ramanathan \& Carmichael \cite{Ramanathan:2008} has focused attention on the role of aerosol black carbon or soot in climate variability, with net positive radiative forcing predicted.  
However, the total black carbon (charcoal + soot) produced as a result of biomass burning may also have a role in long-term carbon sequestration and storage that is significant both quantitatively and in terms of our response to biomass combustion. 

The production of black carbon from biomass burning is governed by the thermal decomposition chemistry of cellulose,  the major constituent of the terrestrial biomass and by far the most abundant biopolymer on earth.   

\begin{figure}[ht]\centerline{
\includegraphics[scale=1]{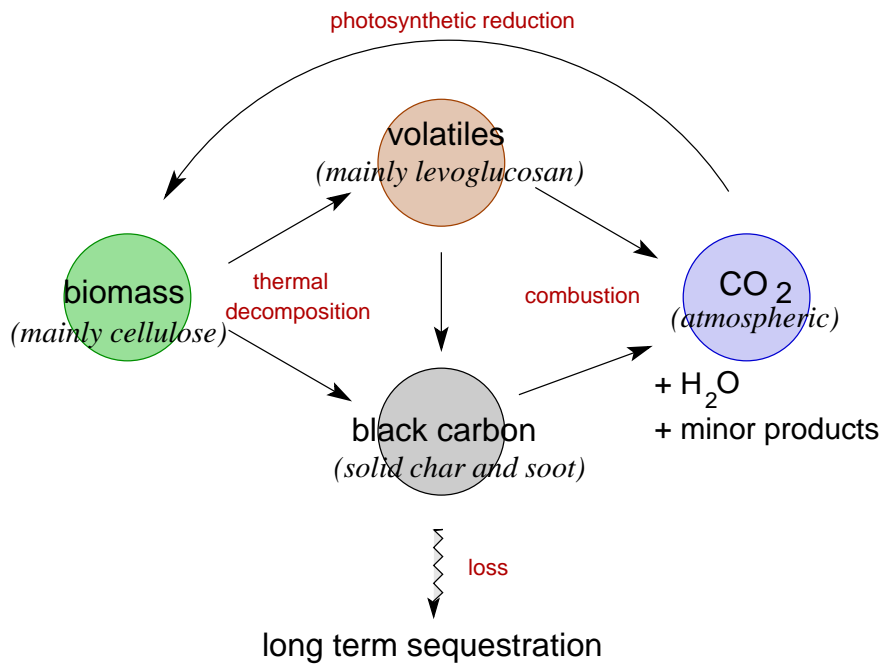}
}
\caption{\label{figure1} In this simplified (but not inaccurate) world carbon accumulates in the long time scale ($>$ 6000 years) reservoir of black carbon which therefore becomes a loss channel, relative to the short time scale (1--200 years) of carbon accumulation in the biomasss and atmospheric CO$_2$ reservoirs. The processes that distribute carbon between flammable volatiles and black carbon are reciprocally linked, or competitive. }
\end{figure}
The schematic in figure \ref{figure1} shows the key carbon reservoirs and transfer channels involved in the biomass burning cycle. Under the thermal stress of a fire cellulose decomposes to yield combustible volatiles (mainly the anhydrosugar levoglucosan) and solid black carbon residue. Crucial to fire ecology is the \textit{competitive} nature of these two processes --- volatiles are produced \textit{at the expense of} char and vice versa. The reciprocally linked formation of volatiles and char was first suggested formally in  \cite{Kilzer:1965} and has been verified by numerous experiments since. A small fraction of the hot volatile gases also condenses and dehydrates to black carbon, or soot, as indicated in the figure. During normal biomass burning conditions the volatiles oxidize to carbon dioxide, water, and other minor products (flaming combustion). Where the temperature is high enough a small amount of the black carbon also burns (glowing combustion), but most  remains intact chemically and joins a stable pool of carbon sequestered for the long term.

\begin{figure}[hb]\centerline{
\includegraphics[scale=1]{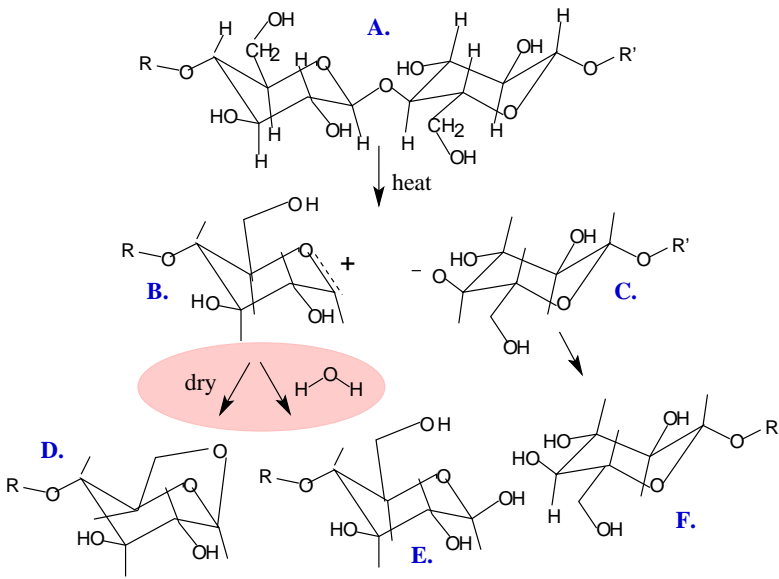}
}
\caption{\label{figure2} Competitive nucleophilic addition chemistry at the heart of cellulose thermal decomposition. R and R' designate the remainders of the cellulose chains on reducing and non-reducing sides of the two-unit section \textbf{A}. To unclutter the sketches the symbols for hydrogen atoms attached to the carbon skeleton are included only for \textbf{A}.}
\end{figure}

Cellulose thermal decomposition chemistry is complicated and not all of the reaction pathways are known, but the competing chemical reactions at the core of the process are exemplars of well-known nucleophilic addition chemistry. They are shown in figure \ref{figure2}. 

The primary structure of cellulose consists of $\beta$-1,4 linked glucopyranose units in alternating orientation in a linear chain, a two-unit section of which is sketched in figure \ref{figure2}, labeled as \textbf{A}. Thermolysis of \textbf{A} yields positively charged \textbf{B} and negatively charged \textbf{C} fragments. The competing reaction steps are highlighted in the figure: the resonance-stabilized positive centre of \textbf{B} may be attacked by the hydroxyl group on C$_6$ of the unit to produce the cyclized levoglucosan end \textbf{D} \textbf{or} by a water molecule to produce a reactive reducing end \textbf{E}. Thermolysis at the next glycosidic linkage of \textbf{D} releases the
volatile levoglucosan. The species \textbf{E} undergoes subsequent dehydration, decarbonylation, aromatization, and cross-linking
reactions that produce black carbon. (The negatively charged fragment \textbf{C} rapidly picks up a positive hydrogen ion to form a relatively unreactive non-reducing end \textbf{F}.) More details and references on the chemistry and thermokinetics of cellulose thermal decomposition can be found in \cite{Ball:1999} and \cite{Ball:2004}. 

The black carbon that is ultimately formed from reducing end fragments \textbf{E} has a polymeric aromatic to graphitic structure that is highly resistant to  chemical, photochemical, and enzyme attack. It is believed to have accumulated in marine sediments over gelogical time, indicating that it is a long-term carbon sink \cite{Kuhlbusch:1995,Preston:2006}, although quantification of black carbon in soils and sediments has large error bars \cite{Hammes:2007}. In the simplified (but not inaccurate) world of Figure \ref{figure1} the only mass loss channel is via the black carbon reservoir and \textit{any non-zero production of black carbon must inevitably result in carbon sequestration}, which could become significant over cycles of burning and regrowth. 

How significant is black carbon as a carbon sink? Some reasonable order-of-magnitude quantitation  would be helpful, at least as a starting point. In the simplest case one assumes that burnt biomass is renewed annually. The amount of carbon stored in the atmosphere is estimated as 750 Gt and the amount in the vegetation reservoir is estimated as 610 Gt \cite{Houghton:1996}. The mass percentage of the carbon in the biomass that is burned each year is taken as as 1\%.  (Wirth et al. \cite{Wirth:2002} estimated that 33\% of net carbon gain in Siberian forests is lost to fire, so this does not seem unreasonable.) The average fraction of biomass carbon converted to black carbon during each burning cycle is taken as 5\%~\cite{Forbes:2006}. (Published estimates range from $\sim$40\% \cite{Seiler:1980} to $\sim$3\% \cite{Fearnside:1993}.) After 100 years of annual burning and regrowth under these conditions the amount of black carbon produced is 30.5 Gt and the mass of carbon in the atmosphere has been reduced by 4.1\%.  

In this simple model the rate of production of black carbon is effectively first order in the amount of carbon in the atmospheric reservoir. The rate or time constant is  $\ll 1$yr$^{-1}$ ($\sim$4$\times 10^{-4}$ yr$^{-1}$) so that over a hundred years the accumulation of black carbon is linear to a good approximation. Of course one does not expect that atmospheric CO$_2$ would asymptote to zero,  disappearing into the black carbon sink through smaller and smaller conflagrations of an ever-diminishing biomass. In reality other carbon pathways may take over. For example, black carbon functions as a source of atmospheric molecular oxygen~\cite{Kuhlbusch:1995}, an increase of which may favour soot reduction by cleaner burning of the volatiles in figure~\ref{figure1}, thus returning more  CO$_2$ to the atmosphere. On the other hand, black carbon is also a source of water vapour. Reference to figure \ref{figure2} indicates that increased water may favour the black carbon pathway in figure \ref{figure1}, having an autocatalytic effect. 

These hypothetical examples fall into the category of ``educated speculations'' at present, but they, and this discussion as a whole, do highlight several issues that were not canvased in Ramanathan's \& Carmichael's \cite{Ramanathan:2008} review: 1) No biomass carbon at all can be sequestered into the stable black carbon sink unless the biomass is burnt regularly; 2)~Cleaner burning of biomass (and fossil) fuels inevitably stores more CO$_2$ in the atmosphere at the expense of black carbon; and 3) The competition between volatilization and charring in figure \ref{figure1} that rules the  thermal decomposition of cellulosic biomass  can be tipped one way or the other by management or engineering practices. 

It is also interesting to ponder that the most prolific biopolymer on earth has exactly the right competitive thermochemical properties to prevent it from swamping the earth, thwart the development of intolerable populations of termites, and provide a stable carbon sink. Its physical chemistry should not be neglected.

\end{document}